\documentclass[useAMS,usenatbib]{mn2e}

\usepackage{amsmath}
\usepackage{amssymb}
\newcommand{{\bfv}}{\mbox{\boldmath$v$\unboldmath}}
\newcommand{{\bfa}}{\mbox{\boldmath$a$\unboldmath}}
\newcommand{{\bfmu}}{\mbox{\boldmath$\mu$\unboldmath}}

\title[The magnetar origin of pulsars]{The magnetar origin of pulsars}
\author[Ricardo Heras]{Ricardo Heras\thanks{E-mail:ricardoherasosorno@gmail.com} \\
Preparatoria Abierta de la SEP en Toluca Edo.~de M\'exico}
\begin{document}


\maketitle

\label{firstpage}

\begin{abstract}
This paper suggests the idea that all neutron stars experienced at birth an ultrafast decay of their magnetic fields from their initial values to their current surface values. If the electromagnetic energy radiated during this field decay is converted into kinetic and rotational energies of the neutron star then the decay time is of the order of $10^{-4}$s provided that the initial magnetic fields lie in the range of $10^{15}\!-\!10^{16}$G and the initial periods in the range $1\!-\!20$ ms. This means that all neutron stars are born with magnetic fields typical of magnetars and periods typical of millisecond pulsars. More explicitly, energy considerations and a birth-ultrafast-magnetic-field decay imply a model that consistently relates birth properties of neutron stars such as the initial period and the initial magnetic field with current properties such as the observed period, the surface magnetic field and the transverse velocity. This model provides a solution to the long-standing problem of finding a physical explanation for the observed high velocities of neutron stars. These stars acquire their space velocities during the birth ultrafast decay of their magnetic fields. The origin of this field decay points to magnetic instabilities, which are inevitable if neutron stars are born as magnetars.
\end{abstract}

\begin{keywords}
Neutron stars, magnetic fields, birth velocities.
\end{keywords}

\section{Introduction}
The existence of millisecond pulsars, radio pulsars and magnetars shows that neutron stars display distinctively different physical properties.
Therefore the idea of a unified model for the diverse classes of neutron stars emerges as an outstanding problem in astrophysics (Kaspi 2010).

The development of a unified picture of neutron stars faces the problem that the diverse families of neutron stars involve some ``irregular" members. For instance,
the neutron stars PSR J2144–-3933 and SGR 0418+5729 have the respective periods $P\!=\!8.5$ s and $P\!=\!9.1$ s, which are typical of magnetars, but they have the respective magnetic fields $B_s\!= \!1.9 \times \!10^{12}$G and $B_s\!=\!7.5 \times \!10^{12}$G, which are typical of radio pulsars (Tiengo et al 2011; Rea et al. 2010).  The neutron star PSR J0537-6910 with $P= 16$ ms and $B_s\!=\!9.25\!\times \!10^{11}$G exhibits a period of a millisecond pulsar but a magnetic field of an ordinary pulsar [unless otherwise specified, all values of periods, magnetic fields and transverse velocities are taken from the ATNF Pulsar Catallogue V 1.43 (www.atnf.csiro.au/research/pulsar/psrcat; Manchester et al. 2005)].
The neutron star PSR J1846-0258  with a period typical of radio pulsars $P= 32$ ms has exhibited magnetar-like X-ray bursts. This pulsar has the magnetic field $B_s\!=\!5\times \!10^{13}$G, which is above the conventional upper limit for radio pulsars but below the traditional lower limit for magnetars.

A second problem for a unified model of neutron stars deals with the existence of neutron stars with similar periods but significantly different magnetic fields or with similar magnetic fields but different periods. A unified model for neutron stars should explain why this can occur. Consider, for example, the set of pulsars: J0700+6418, J0147+5922 and J1917+1353, which have the same period: $P_s=.19$ s but different magnetic fields: $B_s=1.2\times \!10^{10}$G, $B_s=2.3\times \!10^{11}$G and $B_s=1.2\times \!10^{12}$G respectively. Looking for an explication of this result one can see that the respective transverse velocities of these pulsars are: $v_{\perp}=22.8$ km/s, $v_{\perp}=96.9$ km/s and $v_{\perp}=115.8$ km/s. A naive conclusion would be that pulsars with similar periods can  exhibit different magnetic fields because they have different transverse velocities. According to this simplistic conclusion, the magnetic field of the above set of pulsars increases with the transverse velocity. This conclusion seems to be confirmed by noting that the pulsars J0454+5543 and J2022+2854 have the same period: $P_s=.34$ s, similar magnetic fields:
$B_s=9.1\!\times \!10^{11}$G and  $B_s=8.1\!\times \!10^{11}$G, and close transverse velocities (taking into account the uncertain in the determination of these velocities):  $v_{\perp}=317$ km/s and $v_{\perp}=308$ km/s. Nevertheless, one can easily see that the above conclusions are the result of effects of selection.
Consider, for example, the neutron stars J0151-0635 and J2022+2854 which have the same magnetic field: $B_s=8.1\!\times \!10^{11}$G, the same transverse velocity: $v_{\perp}=307$ km/s but very different periods: $P_s=1.4$ s and $P_s=.34$ s. Another example: the pulsars  J1833-0827 and J0835-4510 have the same period $P_s=.08$ s, different magnetic field: $B_s=3.4\!\times \!10^{12}$G and  $B_s=8.9\!\times \!10^{11}$G and distinct transverse velocities: $v_{\perp}=79$ km/s and $v_{\perp}=902$ km/s. From these values one could equally conclude that for the case of pulsars with similar periods, the magnetic field  decreases when the transverse velocity increases.
The idea of having a fundamental relation among current properties of a neutron star such as its period, magnetic field and transverse velocity is then a subtle question. Furthermore, the possible relation between magnetic field and space velocity of neutron stars has been debated for many authors. With respect to the current velocity of neutron stars, there is observational evidence that shows that it was produced by a kick at birth (Spruit \& Phinney, 1998). This suggests that the initial values of period and magnetic field
should be considered in the task of finding a relation among the currents properties of neutron stars. However, little is known about initial periods and magnetic fields of newborn neutron stars. It is natural to think that a ``Grand Unification" of neutron stars should be based on physical first principles, which eventually could connect birth properties of neutron stars such as the initial period $P_0$ and the initial magnetic field $B_0$ with current properties such as the surface magnetic field $B_s$, the actual period $P_s$ and the transverse velocity $v_{\perp}$.

 A first-principles model based on energy conservation and an ltrafast magnetic field decay is presented in this paper, which consistently relates the initial properties: $P_0$ and $B_0$ with the current properties: $P_s, B_s$ and $v_{\perp}$. The general idea is as follows. A neutron star experienced  at birth an abrupt exponential decay of its magnetic field from an initial value $B_0$ to its current surface value $B_s$. As a consequence of this magnetic field decay, the neutron star radiated electromagnetic energy during a characteristic time $\tau_s$. If this radiative energy is converted into kinetic and rotational energies of the neutron star then the time $\tau_s$ is of the order of $10^{-4}$s provided that the initial magnetic field lies in the range of $10^{15}\!-\!10^{16}$G and the initial period in the range $1\!-\!20$ ms. This model supports the idea that most neutron stars are born with magnetic fields of magnetars and periods of millisecond pulsars. The birth magnetic instabilities in neutron stars seem to be the most likely cause for the ultrafast magnetic field decay. The model proposed here provides an explanation for the observed high velocity of neutron stars, according to which the origin of these velocities is in the birth ultrafast decay of their magnetic fields.
The main motivation for this paper is to suggest a basic ingredient for a grand unification of neutron stars.

\section{BIRTH ultrafast magnetic field decay}
The Larmor formula for the power radiated by a time-varying magnetic dipole moment $P=2\ddot{\mu}^2/(3c^3)$  and the estimate
$\ddot{\mu}\!\sim\!\mu_0/\tau^2$, where $\tau$ is the characteristic time in the exponential field decay law $B(t)\!=\!B_0e^{-t/\tau},$ imply the equation
$P\!\simeq \!2\mu_0^2/(3c^3\tau^4),$ which can be used together  with the relation $\mu_0\!=\!B_0R^3/2$ to yield the power radiated by a neutron star of radius $R$ and an initial magnetic field $B_0$:
\begin{align}
P\!\simeq \!\frac{B_0^2 R^6}{6c^3\tau^4}.
\end{align}
This power is restricted by two physical conditions:
\begin{align}
B_0\lesssim 10^{18}{\rm G},\\
\tau\geq R/c.
\end{align}
The condition in Eq.~(2) is imposed by the Virial Theorem of Magnetohydrostatic equilibrium (Lai $\&$ Shapiro 1991).
The condition in Eq.~(3) is required for a physical kick. The time taken for light to cross the pulsar is $R/c\!\approx\!3.33\times\! 10^{-5}$s. Below this value the radiative energy cannot provoke the neutron star motion. From physical considerations, the speed of light $c$ has to be the fastest speed to administer the overall kick to neutron stars, whatever may be the cause of the kick.

Consider now the specific time $\tau_s$ elapsed during the field decay from the initial value $B_0$ to the current surface magnetic field $B_s$. The condition $B(\tau_s)\!=\!B_s$ and the law $B(t)\!=\!B_0e^{-t/\tau}$ imply $B_s\!=\!B_0e^{-\tau_s/\tau}$ or equivalently
\begin{align}
\tau_s\!=\!\tau\ln(B_0/B_s).
\end{align}
From Eqs.~(1) and (4) it follows that the electromagnetic energy radiated $E_{\rm rad}\!\simeq \tau_s P$ is explicitly given by
\begin{align}
E_{\rm rad}\!\simeq \frac{\!B_0^2 R^6\ln(B_0/B_s)}{6c^3\tau^3}.
\end{align}
The initial magnetic field $B_0$ is associated with the initial angular frequency $\Omega_0$ (or equivalently with its associated initial period $P_0$). The initial rotational kinetic energy of a neutron star reads $I\Omega_0^2/2$, where $I=2MR^2/5$ is the inertia moment of the star. The rotational kinetic energy associated with the current surface magnetic field $B_s$ is  $I\Omega_s^2/2$, where $\Omega_s$ is the current angular frequency (and $P_s$ its associated period) of the neutron star. It is expected that the angular frequency of neutron stars decreases during the abrupt field decay of this star. This means that $\Omega_0>\Omega_s$ and so the rotational kinetic energy must decrease. Since $\Omega=2\pi/P$ it follows that the period increases during the birth process ($P_s>P_0$).
By assuming that the radiative energy  in Eq.~(5) is converted into kinetic energy $E_{\rm kin}=Mv^2/2$
and rotational energy $ E_{\rm rot}\!=\!4\pi^2MR^2(P_0^{-2}\!\!-\!\! P_s^{-2})/5$, where $M, R$ and $v$ are the mass, radius and space velocity of the neutron star
then energy conservation reads
\begin{align}
\underbrace{\frac{B_0^2R^6\ln(B_0/B_s)}{6c^3\tau^3}}_{E_{\rm rad}}\!=\! \underbrace{\frac{M v^2}{2}}_{E_{\rm kin}}\!+\! \underbrace{\frac{4\pi^2MR^2}{5}\bigg(\frac{1}{P_0^2}\!-\!\frac{1}{P_s^2}\bigg)}_{E_{\rm rot}},
\end{align}
which implies the following expression for the time $\tau$:
\begin{align}
\tau= \bigg(\frac{5 B_0^2R^6\ln(B_0/B_s)P_s^2P_0^2}{3c^3M(5v^2P_s^2P_0^2-8\pi^2R^2(P_0^2-P_s^2))}\bigg)^{1/3}.
\end{align}
To get an idea of the order of magnitude of $\tau$, consider the Crab pulsar B0531+21 which has $M\!= \!1.4 M_\odot$; $R\!=\!10$ km; $P_s\!=\!0.033$ s$; B_s\!=\!3.78\!\times\!10^{12}$G; $ v_{\perp}\!=\!141$ km/s which implies
$v\!=\!172.7$ km/s [the space velocity $v$ is obtained from the transverse velocity $v_{\perp}$ by means of $v\!\approx\!\sqrt{3/2}\:v_{\perp}$ (Lyne $\&$ Lorimer 1994; Hobbs et al. 2005)]. Without considering a birth ultrafast magnetic field decay, it has been suggested that the Crab pulsar was born with $P_0\!\approx\!.019$ s (Lyne et al 1993). Using this initial period and assuming $B_0\!=\!5.8\times\!10^{15}$G, Eq.~(7) yields
\begin{align}
\tau\! \approx R/c.
\end{align}
It should be noted that if $B_0\!<\!5.8\times\!10^{15}$G then Eq.~(7) gives $\tau\!<\!R/c$, which violates the condition in Eq.~(3). However, one could consider the possibility that $B_0\!>\!5.8\times\!10^{15}$G which implies $\tau\!>\!R/c$. The highest value of the magnetic field allowed by the condition in Eq.~(2) is $B_0\!\sim\!10^{18}$G for which Eq.~(7) yields $\tau\!\sim\!10^{-3}$ s. Therefore, if the values of the initial field $B_0$ of the Crab pulsar are in the interval $5.8\times10^{15}\!\leq B_0\lesssim \!10^{18}$G then the values of the characteristic time $\tau$ are in the interval $R/c\leq \tau\lesssim \!10^{-3}$ s. In any case, it is clear that the small values of the time $\tau$ indicate an ultrafast decay of the magnetic field.

The idea that all neutron stars are born as magnetars is consistent with the choice of the lower limit in Eq.~(2):
\begin{align}
\tau=R/c.
\end{align}
From the example of the Crab pulsar one can see that this choice is associated  with initial magnetic fields in the range $10^{15}\!-\!10^{16}$G, which is the typical range of magnetic fields of magnetars. 
One should note that the choice in Eq.~(9) is consistent for most neutron stars. Consider, for example, the magnetar  J1809-1943 which has $B_s\!=\!2.1\times\!10^{14}$G$; P_s\!=\!5.54$ s and $v=278$ km/s. If the initial values $P_0\!=\!.02$ s and $B_0=\!9.5\!\times\!10^{15}$G are assumed then $\tau\approx R/c$. Finally, consider the isolated millisecond pulsar B1257+12 which has $B_s\!=\!8.53\times\!10^{8}$G$; P_s\!=\!.006$ s and $v=429$ km/s. It has been suggested that this millisecond pulsar was born with approximately its current period (Miller \& Hamilton, 2001). Under this assumption, Eq.~(7) with $P_0=P_s$ gives again the value $\tau\!\approx R/c$ provided that the initial field $B_0=\!1.05\!\times\!10^{15}$G is assumed. Equations~(4) and (9) yield
$\tau_s\!=\!(R/c)\ln(B_0/B_s).$  For magnetic field decays ranging from one to eight orders of magnitude it follows that $2.3  \leq\ln(B_0/B_s)\leq4.4$. Therefore
\begin{align}
\tau_s\!\sim\!10^{-4}\; {\rm s}.
\end{align}
This means that the decay from $B_0$ to $B_s$ is ultrafast if the initial magnetic fields lie in the range of $10^{15}\!\!-\!\!10^{16}$G and the initial periods in the range $1-20$ ms, that is, if all neutron stars are born with magnetic fields typical of magnetars and periods typical of millisecond pulsars.

\section{Initial magnetic fields of neutron stars}
Using Eq.~(9), the energy conversion in Eq.~(6) reduces to
\begin{align}
\frac{B_0^2R^3\ln(B_0/B_s)}{6}=\frac{M v^2}{2} + \frac{4\pi^2MR^2}{5}\bigg(\frac{1}{P_0^2}-\frac{1}{P_s^2}\bigg).
\end{align}
This fundamental equation involves five known (current) parameters: $P_s, B_s, M, R$ and $v$. However, Eq.~(11) also involves two unknown birth (initial) parameters: $P_0$ and $B_0$. The simplest way to work with Eq.~(11) consists in assuming the value of either $P_0$ or $B_0$. The value of the remaining parameter will be then determined.

From Eq.~(11) one can obtain the initial magnetic field
\begin{align}
B_0= B_s e^{W([6Mv^2+\:(8/5)\pi^2MR^{2}(P_0^{-2}-P_s^{-2})]B_s^{-2}R^{-3})/2},
\end{align}
where $W(x)$ is the Lambert function, which is defined as the inverse of the function $f(x)\!=\!xe^{x}$ satisfying $W(x)e^{W(x)}\!=\!x$.
If one writes $v\!\approx\!\sqrt{3/2}\:v_{\perp}$, $M\!= \!1.4 M_\odot$ and $R\!=\!10$ km into Eq.~(12) then  it reduces to
\begin{align}
B_0= B_s e^{W([\sigma v_{\perp}^2+\:\lambda(P_0^{-2}-P_s^{-2})]B_s^{-2})/2},
\end{align}
where $\sigma\!=\!2.52\! \times \!10^{16}$gr cm$^{-3};\lambda\!\approx \!2.65\times \!10^{29}$G$^{-2}$ s$^{-2}$.
To apply Eq.~(13) one needs to consider neutron stars whose $P_s, B_s$ and $v_{\perp}$ are known. In Eq.~(13) the value of $P_0$ should be assumed. It is generally believed that neutron stars are born with initial periods of order of milliseconds. This belief allows one to assume initial periods of neutron stars in the range $1\!-\!20$ s. The following sets of neutron stars will be now considered.

Interval: $2$ s $\!<\!P_s\!\leq\!8.5$ s. There are 9 neutron stars in this interval. The average values of this set are $\widetilde{P_s}=4.82$ s, $\widetilde{B_s}=3.14\times 10^{13}$G and $\widetilde{v_{\perp}}=226.55$ km/s.
If one assumes $\widetilde{P_0}=.02$ s then Eq.~(13) predicts ${B_0}=7.8\times10^{15}$G. Some particular neutron stars of this set deserve to be mentioned. The magnetar  J1809-1943 is characterized by the values $B_s\!=\!2.1\times\!10^{14}$G$, P_s\!=\!5.54$ s and $v_{\perp}=227$ km/s (Ibrahim et al. 2004; Helfand et al. 2007). If the initial period $P_0\!=\!.02$ s is assumed then Eq.~(13) predicts that this magnetar was born with the magnetic field $B_0\!\approx \! 9.4\times \!10^{15}$G. The magnetar PSR J1550-5418 has
$B_s\!=\!2.22\times\!10^{14}$G and $P_s\!=\!2.06$ s. It has recently been reported that this magnetar has $v_{\perp}\!=\!260$ km/s (Deller et al. 2012).
If the initial period $P_0\!=\!.02$ s is assumed then Eq.~(13) predicts that this magnetar was born with the magnetic field $B_0\!\approx \! 9.5\times \!10^{15}$G. This means that
the magnetars J1809-1943 and J1550-5418 continue being magnetars after their birth.
As already mentioned, the enigmatic pulsar PSR J2144–-3933 with period $P_s\!=\!8.5$ s exhibits magnetar properties but its magnetic field is typical of radio pulsars: $B_s\!=\!1.9\times\! 10^{12}$G. It has recently been reported  that this pulsar has $v_{\perp}\!=\!135$ km/s (Tiengo et al. 2011). Taking into account these values for $B_s, P_s, v_t$ and assuming $P_0\!=\!.02$ s, Eq.~(13) predicts that the initial magnetic field of this pulsar was $B_0\!\approx \! 6.41\times \!10^{15}$G indicating that this pulsar was born as a magnetar.

Interval: $1$ s$\leq P_s\leq 2$ s. There are 29 pulsars in this set of neutron stars. The average values are $\widetilde{P_s}=1.32$ s, $\widetilde{B_s}=4.17\times 10^{12}$ G and $\widetilde{v_{\perp}}=261.55$ km/s.
If $\widetilde{P_0}=.02$ s then $B_0=6.78\times 10^{15}$G.

Interval: $.02$ s$ \leq P_s< 1$ s.
There are 130 neutron stars with periods in this interval. The corresponding average values are $\widetilde{P_s}=0.41$ s, $\widetilde{B_s}=1.28\times 10^{12}$G and $\widetilde{v_{\perp}}=409.06$ km/s. Equation~(13) predicts that if $\widetilde{P_0}\!=\!.02$ s then $B_0=6.42\times 10^{15}$G. According to Eq.~(13), the Crab pulsar  B0531+21 with $v_{\perp}\!=\!141$ km/s, $P_s=.033$ s and $B_s\!=\!3.78\!\times\!10^{12}$G was born with $B_0\!\approx\!5.8\!\times\!10^{15}$G if $P_0=.019$ s. The vela pulsar B0833$-$45 with  $v_{\perp}\!=\!78$ km/s, $P_s=.089$ s and $B_s\!=\!3.38\!\times\!10^{12}$G was born with $B_0\!\approx\!6.4\!\times\!10^{15}$G if the initial period  $P_0\!=\!.02$ s is assumed. For the Guitar pulsar PSR B2225+65 with $v_{\perp}\!=\!1729$ km/s, $B_s\!=\!2.6\!\times\!10^{12}$G and $P_s=.6$ s, Eq.~(13) states that this pulsar was born with $B_0\!\approx\!9.3\times10^{15}$G if the initial period  $P_0\!=\!.02$ s is assumed.

Isolated millisecond pulsars in the interval: $.0015$ s $\! \leq P_s<\! .009$ s. There are 9 pulsars in this set of neutron stars. The average values are $\widetilde{P_s}\!=\!0.005$ s, $\widetilde{B_s}=2.59\times 10^{8}$G and $\widetilde{v_{\perp}}\!=\!60.11$ km/s. One can consider a very small change in the period by assuming, for example, the average initial period $\widetilde{P_0}\!=\!.0049$ s. Under this assumption Eq.~(13) predicts $B_0\!=\!3.65\times 10^{15}$G. If one considers a considerable change in the period then
the initial magnetic field also considerably increases. If, for example, $\widetilde{P_0}\!=\!.001$ s then $B_0\!=\!8\times\!10^{16}$G which is very unlikely.
 Therefore, no significant change of the initial of isolated millisecond pulsars is expected during the birth process. For the isolated millisecond pulsar B1937+21 with $v_{\perp}\!=\!31.78$ km/s, $B_s\!=\!4.09\!\times\!10^{8}$G and $P_s=.00155$ s, Eq.~(13) predicts $B_0\!=\!6.58\times 10^{15}$G if $P_0\!=\!.00154$ s is assumed.

It is clear that all neutron stars are born with initial magnetic fields
in the range of $10^{15}\!-\!10^{16}$G provided that the associated initial periods lie in the range $1\!-\!20$ ms.

\section{Initial periods of neutron stars}

Equation~(11) implies the following formula for the initial period of neutron stars:
\begin{equation}
P_0\!=\! \sqrt{\frac{24M\pi^2R^2P_s^2}{5P_s^2(B_0^{2}R^3\ln(B_0/B_s)\!-\!3Mv^2)\!+\!24\pi^2MR^2}}\;.
\end{equation}
Having in mind the idea that neutron stars are born as magnetars, one can assume the ``canonical'' value $B_0=10^{16}$G. Considering also the usual canonical values $M\!= \!1.4 M_\odot$ and $R\!=\!10$ km it follows that Eq.~(14), applied to the four sets of neutron stars discussed in the preceding section, predicts that the average initial periods of these sets of stars are in the range $4\!-\!15$ ms. In fact, consider again the set of neutron stars in the interval $2$ s $\!<\!P_s\!\leq\!8.5$ s the average values are $\widetilde{P_s}=4.82$ s, $\widetilde{B_s}=3.14\times 10^{13}$G and $\widetilde{v_{\perp}}=226.55$ km/s.
If one assumes $B_0=10^{16}$G then  Eq.~(14) predicts $P_0=.015$ s. For neutron stars with $1$ s$\leq P_s\leq 2$ s the average values are $\widetilde{P_s}=1.32$ s; $\widetilde{B_s}=4.17\times 10^{12}$ G and $\widetilde{v_{\perp}}=261.55$ km/s.
If $B_0=10^{16}$ G then Eq.~(14) gives $P_0=.013$ s. For the subset of neutron stars with $.02$ s$ \leq P_s< 1$ s the corresponding average values are $\widetilde{P_s}=0.41$ s; $\widetilde{B_s}=1.28\times 10^{12}$G and $\widetilde{v_{\perp}}=409.06$ km/s.
Equation~(14) predicts that if $B_0=10^{16}$G then $P_0\!=\!.012$ s. For the case of the isolated millisecond pulsars with $.0015$ s $\! \leq P_s<\! .009$ s the average values are $\widetilde{P_s}\!=\!0.005$ s; $\widetilde{B_s}=2.59\times 10^{8}$G and $\widetilde{v_{\perp}}\!=\!60.11$ km/s. If one assumes the initial magnetic field  $B_0=10^{16}$G then $P_0\!=\!.004$ s.

 \section{Birth velocities of neutron stars}
From Eq.~(11) it follows that
 \begin{equation}
v= \sqrt{\frac{B_0^2R^3\ln{(B_0/B_s)}}{3M}-\frac{8\pi^2 R^2}{5}\bigg(\frac{1}{P_0^2}-\frac{1}{P_s^2}\bigg)},
\end{equation}
According to this equation the birth velocity of neutron stars depends on the initial values of $B_0$ and $P_0$ as well as on the current values of $B_s$ and $P_s$. Equation (15) is highly sensible to small changes in the values of $B_0, P_0, B_s$ and $P_s$.

As an application of Eq.~(15) consider the recently discovered SGR 0418+5729, which is similar to other magnetars in all observed properties except for the fact that it has a surface magnetic field typical of radio pulsars: $B_s\!\simeq \!7.5\!\times \!10^{12}$G (Rea et al. 2010). The transverse velocity of this pulsar has not been reported yet. The current period of SGR 0418+5729 is $P_s=9.1$ s. By assuming that the initial period of this pulsar was in the range $.015-.15$ s, Eq.~(15) yields a real velocity $v>0$ if the initial magnetic field was in the range $10^{15}\!-\!10^{16}$G, which allows one to conclude that the SGR 0418+5729 was born as a magnetar.

 If one writes $v\!\approx\!\sqrt{3/2}\:v_{\perp}$, $M\!= \!1.4 M_\odot$ and $R\!=\!10$ km into Eq.~(15) then it yields the following formula to calculate the transverse velocity
 \begin{equation}
v_{\perp}= \sqrt{k_1B_0^2\ln{(B_0/B_s)}\!-\!k_2(P_0^{-2}\!-\!P_s^{-2})},
\end{equation}
 where the constants $k_1\!=\!7.9365\!\times \!10^{-17}$cm$^2$s$^{-2}$G$^{-2}$ and $k_2\!=\!1.0528\!\times \!10^{-13}$cm$^2$ have been introduced. Equation~(16) states that the birth transverse velocity of a neutron star is simultaneously related with two initial properties: $P_0$ and $B_0$  and two current properties: $P_s$ and $B_s$.
 Previous attempts to relate  the transverse velocity $ v_{\perp}$ with the surface magnetic field $B_s$ were unsatisfactory because they did not involve the values of
 $B_0, P_0$ and $P_s$. Equation~(16) states that if a typical pulsar with $B_s\!=\!10^{12}$G and $P_s=.5$ s was born with $B_0\!=6.308\times\!10^{15}$G and $P_0=.02$ s
 then the birth sudden field decay (from $B_0$ to $B_s$) produces the transverse velocity $ v_{\perp}\!=\!367.8$ km/s which implies the space velocity $ v\!=\!450$ km/s. This is the average velocity for pulsars suggested by Lyne and Lorimer (1994).

For the case of isolated millisecond pulsars the values of the transverse velocity given by Eq.~(16) are too sensitive to changes of the initial period and the initial  magnetic field. If $B_0\!=\!10^{16}$G is taken to be the canonical value for initial magnetic field of isolated millisecond pulsars with average values: $\widetilde{P_s}\!=\!0.005$ s and $\widetilde{B_s}=2.59\times 10^{8}$G then Eq.~(16) gives the transverse velocity $v_{\perp}\!=\!60.14$ km/s when the initial period  $P_0\!=\!0.004336955$ is assumed. But if one assumes the slightly different value $P_0\!=\!0.004338$ s, for example, then Eq.~(16) yields the velocity $v_{\perp}\!=\!174.87$ km/s.

\section{Discussion}
The energy conversion in Eq.~(11) is the fundamental formula for the model proposed here. It links birth properties of neutron stars with the current properties of these stars.
Equation (11) in its diverse expressions given in Eqs.~(12)-(16) can be considered as a basic ingredient for a grand unification of the neutron star families.

The irregular properties of some members of the pulsar families can be justified to a certain extent by Eq.~(11). For example, the PSR J2144–-3933 with period of a magnetar: $P_s\!=\!8.5$ s and magnetic field of radio pulsar: $B_s\!=\!1.9\times\! 10^{12}$G does not constitute an enigma for the model based on Eq.~(11). This pulsar has the transverse velocity $v_{\perp}\!=\!135$ km/s. Consistence of Eq.~(11) suggests  that this pulsar could have be born with the initial period $P_0=.02$ s and the initial magnetic field $B_0\!=\!6.41\times \!10^{15}$G. During the birth process the initial magnetic field of this neutron star decayed about three order of magnitude and its initial period increased about two orders of magnitude. This example illustrates the general prediction of Eq.~(11) that neutron stars are born with magnetic fields of magnetars and periods of millisecond pulsars. On the other hand, Eq.~(11) mixes the values of the birth and current properties in such a way that one may have neutron stars with similar periods but significantly different magnetic fields or with similar magnetic fields but different periods.

A physically viable explanation for the assumed birth ultrafast magnetic field decay could be found, in principle, by considering magnetic instabilities experienced by newborn neutron stars. Geppert $\&$ Rheinhardt (2006) have discussed a magnetohydrodynamical process (MHD) that significantly reduces the initial magnetic field of a newly-born neutron star in fraccions of a second. Such a reduction is due to MHD-instabilities which evolve on the timescale of the Alfven-time. For magnetic fields in the range of $10^{15}\!-\!10^{16}$G the Alfven times $T_A$ are of the order $10^{-2}\!-\!10^{-3}$s. On the other hand, the decay time $\tau_s$ (from $B_0$ to $B_s$) of the model proposed here is of the order of $10^{-4}$s. Therefore, the birth exponential decay of the initial magnetic field occurs during a fraction of the time elapsed by MHD-instabilities, that is, the exponential decay occurs in the time interval ranging from $T_A/100$ to $T_A/10$. In other words: the initiation and the end of the decaying of the magnetic field occur during the magnetic instabilities of newly-born neutron stars. It should also be noted that during the extremely brief times of the exponential decay of newborn neutron stars there is no convection and the electric conductivity is constant in time and space. The assumption of the exponential decay could also be justified a posteriori by noting that it allows one to formulate the unified model of neutron stars presented here, which is consistent with observations provided that the initial magnetic fields lie in the range of magnetars and the initial periods in the range of millisecond pulsars.

 It is pertinent to mention that Spruit (2008) has suggested that a differential rotation in the final stages of the core collapse process can produce magnetic fields typical of magnetars. As above emphasized, initial fields in the range of magnetars are required as an initial input for the model proposed here. Spruit argues that some form of magnetorotational instability may be the cause of an exponential growth of the magnetic field during the end of the core collapse and notes that ``Once formed in core collapse, the magnetic field is in danger of decaying again by magnetic instabilities.'' This field decay produced by magnetic instabilities is the basic assumption in this paper.
 According to the model proposed here, at the end of the birth of a neutron star there is a process that can convert all radiative energy produced by an ultrafast field decay, caused in turn by magnetic instabilities, into kinetic and rotational energies of the neutron star. However, it can be argued that one equally could consider that the rotational energy produced in the birth process is transformed into kinetic and radiative energies, that is, that the birth rotation ultimately causes the increase of the space velocity and the loss of radiation. This last interpretation is similar to that of the electromagnetic rocket model of neutron stars proposed by Harrison and Tademaru (1975). They have pointed out that ``...the loss of radiation and increase in translational kinetic energy are both at the expense of the rotational energy.'' Harrison and Tademaru model considers the asymmetric radiation from an off-centered magnetic dipole. Nevertheless, the idea of establishing cause-effect relations among kinetic, rotational and radiative energies of newly-born neutron stars is somewhat subtle mainly because the involved energy transformations occur in extremely short times. In the physical interpretation suggested here, the magnetic instabilities at the end of the birth process produce an ultrafast magnetic field decay, which in turn originates the current periods and proper motions of neutron stars.

\vskip 10pt
\noindent{\bf ACKNOWLEDGMENTS}

\vskip 5pt
I thank professors U. Geppert, E. Parker, D. Lorimer and A. Hewish for  illuminating correspondence.
I am indebted with professor Jos\'e A. Heras for some of his decisive suggestions developed in this paper.

\label{lastpage}

\end{document}